# Propagation of Nonlinear Waves in Multi-component Pair Plasmas and Electron-Positron-Ion Plasmas

Tanvir I. Rajib
*Department of Physics, Jahangirnagar University, Savar, Dhaka-1342, Bangladesh*
*E-mail: tirajibphys@juniv.edu OR tirajibphys@gmail.com*

The propagation of small amplitude stationary profile nonlinear solitary waves in a pair plasma is investigated employing the reductive perturbation technique via well-known Kortewegde Vries (KdV) and modified KdV (mKdV) equations. We tend to derive the exact form of nonlinear solutions, and study their characteristics. Two distinct pair ion species are considered of opposite polarity and same mass, additionally to a massive charged background species, that is assumed to be stationary, given the frequency scale of interest within the pair-ion context, the third species is thought of as a background defect (e.g. charged dust) component. On the opposite hand, the model conjointly applies formally to electron-positron-ion (e-p-i) plasmas, if one neglects electron-positron annihilation. A parametric analysis is carried out, as regards the impact of the dusty plasma composition (background number density), species temperature(s) and background species. It is seen that distinguishable solitary profiles are observed for KdV and mKdV equations. The results are of connection in pair-ion (fullerene) experiments and additionally potentially in astrophysical environments, e.g. in pulsars.

## I. INTRODUCTION

A significant attempt has recently been devoted to pair plasmas (p.p.), a term denoting large ensembles of charged matter consisting of charged particle populations bearing equal masses and opposite charge signs [1–4]. In distinction to standard (electron-ion, e-i) plasmas, wherever the large mass inequality between plasma constituents imposes distinct frequency scales, the combine species (of equal however opposite charge) respond on identical scale. Roughly expectedly, plasma wave characteristics cannot always be deduced from known results for e-i plasmas by merely taking a proper limit of equal masses. As an example, parallel propagating linear electromagnetic (EM) waves are not circularly however linearly polarized in pair plasmas, and Faraday rotation [5] is remarkably absent in this [6]. Moreover, ion-acoustic waves do not have any counterpart in electron-positron (e-p) plasmas, wherever electrostatic wave dispersion might bear high frequency (Langmuir-type) characteristics [3, 4]. Remarkably, the production of a pair fullerene-ion plasmas within the laboratory [7–9] have enabled experimental studies of pair plasmas permitting one to induce rid of intrinsic issues concerned in e-p plasmas, specifically, pair recombination (annihilation) processes.

In general, electron-positron plasmas can also be characterised via way of means of the presence of positive ions, similarly to electrons and positrons. Electron-positron-ion (e-p-i) plasmas occur in different astrophysical contexts, inclusive of the early Universe [10], active galactic nuclei (AGN) [11] and in pulsar magnetospheres [12], and feature additionally been created in the laboratory [13–16]. The standard description of e-p-i plasmas adopted here models them as absolutely ionizing plasmas with two populations of various charge signs possessing equal lots and absolute charge values $(m_+ = m_- = m, q_+ = -q_- = Ze)$, in addition to a population of positively charged ions, with $m_3 = M >> m$ and $q_3 = +Ze$, where $e$ is the magnitude of the electron charge. On the opposite hand, one might anticipate the existence (intrinsically, or by intentional injection/doping) of a little fraction of charged massive particles (a heavier ion species, or dust particulates, defect) into fullerene pair-ion plasma [7–9] so as to understand three-component plasmas which can accommodate new physical phenomena. According to these concerns, we have a tendency to shall henceforward keep the charge sign $s = q_3/|q_3|$ arbitrary in the considered model.

Experimental investigations of low-amplitude (linear) electrostatic (ES) oscillations recommend the existence of three distinct modes [7–9]. Two of those modes, namely an acoustic mode and a Langmuir-like high-frequency mode, are foreseen by theory [3, 4, 17, 18]. An intermediate-frequency mode additionally reportable [7–9] continues to be a subject of debatable discussion among theoreticians, and varied different interpretations are equipped, in terms of soliton trains [19], ion-acoustic waves accelerated by surplus electrons [20] or BernsteinGreeneKruskal (BGK)-like trapped ion modes [21] although the experiments mentioned higher than place confidence in a radially symmetrical pair-component plasma preparation (pure p.p., i.e. equal number densities, and equal temperatures among the pair species), it is tended to choose to leave the density and temperature ratio(s) of the positive-to-negative ion species arbitrary, i.e. not essentially equal to unity (viz. pure p.p.).

Danehkar [22] has studied the electrostatic solitary waves in an electron-positron pair plasma with suprathermal electrons. El-Wakil *et al.* [23] studied the propagation of solitary waves and double-layers in electron-positron pair Plasmas with stationary ions and nonextensive electrons based on reductive perturbation theory via different nonlinear KdV, mKdV and Gardners equations. Dissipative ion acoustic solitary waves in col-lisional, magneto-rotating, non-thermal electronpositronion plasma has been studied by Farooq *et al.* [24]. Nonlinear electrostatic waves are studied in unmagnetized, dissipative pair-ion plasmas in the presence of weak transverse perturbations by Masood *et al.* [25], while the propagation of magnetoacoustic (fast magnetohydrodynamic) waves in pair-ion (PI) fullerene plasma is studied in the linear and nonlinear regimes using reductive perturbation technique (RPT), the Kortewegde Vries (KdV) equation is derived [26]. Several theoretical studies of linear, as well as nonlinear wave phenomena in both unmagnetized and magnetized e-p-i plasmas, have been of interest [27–31], Abdelsalam *et al.* [32] and Rasheed *et al.* [33] investigated ion-acoustic waves in e-p-i plasmas by using different forms of distributions for electrons and positrons, such as the Thomas-Fermi approximation and non-relativistic electron/positron equation of state and observed that the ion-acoustic speed in the degenerate pair-ion plasma depends both on the electron temperature and ion mass as well as the concentration of the positron and ions in the plasma. Alinejad and Mamun [28] studied linear and nonlinear propagation of ion-acoustic waves subjected to an external magnetic field in an e-p-i plasma and found that when the positron concentration is increased, the frequencies of both the slow and fast modes decreased. They investigated the effects of obliqueness, magnetic field strength, densities, and the temperature ratio on such plasmas. For most of these investigations, the general dispersion relation for multi-component plasmas had been obtained without considering the mass of the ions. It is thus of interest to investigate the effects of the ion mass on the structures of the waves propagating in an e-p-i plasma.

In the present work, we are interested in investigating the nonlinear propagation of solitary waves in a three-component pair plasma of either the doped pair plasma or the e-p-i plasma by using both KdV and mKdV equations with reductive perturbation technique. The layout of the manuscript is as follows: The governing equations describing our plasma model are presented in Section II. A brief description of the mathematical technique, followed by the derivation of a KdV and mKdV equations are given in Sections III and IV, respectively. A brief description of results and discussion is presented in Section V and finally, a conclusion is drawn in Section VI.

## II. BASIC EQUATIONS GOVERNING THE PLASMA DYNAMICS

We consider the propagation of electrostatic solitary waves in a magnetized, collision free three components plasma system consisting of positive ions (charge $+Ze$ and mass $m$, where $Z$ is the number of protons residing onto the ions surface and $e$ is the magnitude of the charge of an electron), negative ions (charge $-Ze$ and mass $m$) describe by the fluid-momentum equations, and a third background species of mass $m_3$ and charge $s_3 Z_3 e = \pm Z_3 e$. Here, $s_{j=+,-,3}$ denotes the sign of species j (i.e., $s_+ = -s_- = +1$ and $s_3 = s = \pm 1$). It can be noted that $Z = 1$ and $s = +1$ applies to e-p-i plasmas as well as $Z = 1$ and $s = \pm 1$ in pair-ion (fullerene). Now, the dynamics of such a plasma system can be described by the set of equations as follows

$$\partial_t n_\pm + \boldsymbol{\nabla} \cdot (n_\pm u_\pm) = 0 \qquad (1)$$

$$m(\partial_t + u_\pm \cdot \boldsymbol{\nabla}) u_\pm = \mp Ze \boldsymbol{\nabla} \phi - \frac{1}{n_\pm} \boldsymbol{\nabla} p_\pm$$
$$\pm \frac{Ze}{c}(u_\pm \times \mathbf{B}), \qquad (2)$$

$$\nabla^2 \phi = 4\pi e [Z(n_- n_+) - s_3 Z_3 n_3], \qquad (3)$$

where $n_\pm$ is the number densities and the fluid velocity variables $u_\pm$ and the electrostatic potential $\phi$, while $c$ is the speed of light. The external magnetic field $\mathbf{B}$ is considered to lie along the $x$-axis, i.e., $\mathbf{B} = B_0 \hat{\mathbf{x}}$ (here, $B_0$ is the magnitude of the ambient magnetic field and $\hat{\mathbf{x}}$ is a unit vector along the $x$-axis). The system is closed by the equation(s) of state $p_\alpha \sim n_\alpha^\gamma$ with $\gamma = (N+2)/N$, where $N$ is the degree of freedom. We consider $N = 2$ because of analytical tractability and physical insight.

Now, the normalized equations we get from equations (1)-(3) for positive ion fluid (+ index, $\phi$=+ve) and negative ion fluid (- index, $\phi$=-ve, and $\sigma$=1) as follows

$$\frac{\partial \tilde{n}_\pm}{\partial \tilde{t}} + \frac{\partial \tilde{n}_\pm \tilde{u}_{\pm,x}}{\partial \tilde{x}} + \frac{\partial \tilde{n}_\pm \tilde{u}_{\pm,y}}{\partial \tilde{y}} = 0, \qquad (4)$$

$$\frac{\partial \tilde{u}_{\pm,x}}{\partial \tilde{t}} + \left(\tilde{u}_{\pm,x}\frac{\partial}{\partial \tilde{x}} + \tilde{u}_{\pm,y}\frac{\partial}{\partial \tilde{y}}\right)\tilde{u}_{\pm,x}$$
$$\pm \frac{\partial \tilde{\phi}}{\partial \tilde{x}} + 2\sigma \frac{\partial \tilde{n}_\pm}{\partial \tilde{x}} = 0, \qquad (5)$$

$$\frac{\partial \tilde{u}_{\pm,y}}{\partial \tilde{t}} + \left(\tilde{u}_{\pm,x}\frac{\partial}{\partial \tilde{x}} + \tilde{u}_{\pm,y}\frac{\partial}{\partial \tilde{y}}\right)\tilde{u}_{\pm,y}$$
$$\pm \frac{\partial \tilde{\phi}}{\partial \tilde{y}} + 2\sigma \frac{\partial \tilde{n}_\pm}{\partial \tilde{y}} = 0, \qquad (6)$$

$$\frac{\partial \tilde{u}_{\pm,z}}{\partial \tilde{t}} + \left(\tilde{u}_{\pm,x}\frac{\partial}{\partial \tilde{x}} + \tilde{u}_{\pm,y}\frac{\partial}{\partial \tilde{y}}\right)\tilde{u}_{\pm,z} = 0, \qquad (7)$$

$$\tilde{\nabla}^2 \tilde{\phi} = \tilde{n}_- - \tilde{n}_+ - s\frac{Z_3}{Z}\tilde{n}_3. \qquad (8)$$

Here, the density of the heavy plasma component 3 is taken to be stationary (of fixed density) which implies that the heavy background species will react extremely slowly to variations of the electric potential due to the fast ion dynamics so that static equilibrium (for species 3) can be maintained at all times. Thus, the density $n_{j=+,-,3}$ is normalized by the unperturbed negative ion density $n_0$, $\mathbf{u}_\alpha$ is scaled by the negative ion thermal speed $C_{s-} = \lambda_{D-}\omega_{p-} = \sqrt{(k_B T_-)/m}$ with the negative ion plasma period $\omega_{p-} = \sqrt{(4\pi Z^2 e^2 n_0/m)}$ and the potential $\phi$ by $\phi_0 = k_B T_-/Ze$. The space and time variables are scaled by the negative ion Debye radius $\lambda_{D-} = \sqrt{k_B T_i / 4\pi Z^2 e^2 n_0}$ and the plasma period $\omega_{p-}^{-1}$, respectively. The temperature ratio is equal to positive ion fluid temperature to negative ion fluid temperature i.e., $\sigma = T_+/T_-$. Overall, the charge neutrality condition for our plasma model can be written as

$$\delta + \beta = 1, \tag{9}$$

where $\delta = n_{+,0}/n_0$ (here "0" means unperturbed density states) and $\beta = s\frac{Z_3}{Z}\frac{\tilde{n}_3}{n_0}$. If there is no background species, then $\delta = 1$ and $\beta = 0$ that recovers the pure pair-plasma limit with plasma frequencies of both ion species coincide. For the presence of background species $\delta \neq 0$ (i.e., $\beta \neq 0$). For negative background species $\delta > 1$ and $\beta < 0$, while positive background shows when $0 < \delta < 1$ and $\beta > 0$. Retain that $\beta = 1 - \delta < 1$ by definition, although no lower boundary exists, for a negative species ($s = -1$).

## III. DERIVATION OF THE K-DV EQUATION AND ITS SOLUTION

To study the propagation of small amplitude stationary profile nonlinear electrostatic solitary waves in a pair plasma, we want to derive the Kortewegde Vries equation (K-dV equation) by employing the reductive perturbation theory and for that case, first, we can write the stretched co-ordinates in the form [34]

$$\xi = \epsilon^{1/2}(x - v_p t), \tag{10}$$
$$\tau = \epsilon^{3/2} t, \tag{11}$$

where $v_p$ is the phase speed and $\epsilon$ ($0 < \epsilon < 1$) is a smallness parameter measuring the weakness of the dispersion. Then we can write the dependent variables [35] as

$$\Pi(x,t) = \Pi^{(0)} + \sum_{m=1}^{\infty} \epsilon^m \sum_{m=1}^{\infty} \Pi^{(m)}(\xi,\tau)\,\mathrm{e}^{i(kx-\omega t)}, \tag{12}$$

where $\Pi^{(m)}$ is the m-th order to any among state variables $[n_\pm, u_{\pm,x}, \phi]$, at equilibrium $\Pi^0 = [1, \delta, 0, 0, 0]^T$, and $k$ ($\omega$) is real variables representing the carrier wave number (frequency). The derivative operators can be written as

$$\partial_x \to \epsilon^{1/2}\partial_\xi, \tag{13}$$
$$\partial_t \to \epsilon^{3/2}\partial_\tau - \epsilon^{1/2} v_p \partial_\xi. \tag{14}$$

The transverse velocity (y- and z-components) is assumed to vary on a slower scale, hence:

$$u_{-,y} = \epsilon^{3/2} u_{-,y}^{(1)} + \epsilon^2 u_{-,y}^{(2)} + \epsilon^{5/2} u_{-,y}^{(3)} + \cdots, \tag{15}$$
$$u_{+,z} = \epsilon^{3/2} u_{+,z}^{(1)} + \epsilon^2 u_{+,z}^{(2)} + \epsilon^{5/2} u_{+,z}^{(3)} + \cdots. \tag{16}$$

Now, by substituting (15) and (16) into (4)-(8), and collecting the terms containing $\epsilon$, the first order [$\psi =$ $\phi^{(1)}$] reduced equations can be written as

$$n_-^{(1)} = -(v_p^2 - 2)^{-1}\psi, \tag{17}$$
$$u_{-,x}^{(1)} = -v_p(v_p^2 - 2)^{-1}\psi, \tag{18}$$
$$n_+^{(1)} = \delta(v_p^2 - 2\delta\sigma)^{-1}\psi, \tag{19}$$
$$u_{+,x}^{(1)} = v_p(v_p^2 - 2\delta\sigma)^{-1}\psi. \tag{20}$$

The Poisson equation generates the following relation

$$\frac{1}{v_p^2 - 2} + \frac{\delta}{v_p^2 - 2\delta\sigma} = 0, \tag{21}$$

which determine the phase speed $v_p$ as follows

$$v_p = \sqrt{\frac{(1-\beta)(\sigma+1)}{1-\beta/2}}. \tag{22}$$

The solution of $v_p$ is provided that $v_p \neq \pm\sqrt{2}, \pm\sqrt{2\delta\sigma}$ which excludes propagation in temperature-symmetric electron-positron plasmas, viz. $\delta = \sigma = 1$. $\beta$ determines the role of the background species and $\sigma$ indicates the pair-ion asymmetry.

Eliminating the second order perturbed quantities and making use of the first order results, we obtain a nonlinear partial-derivative equation known as Kortewegde Vries (K-dV) equation in the following form

$$\partial_\tau \psi + A\psi\partial_\xi \psi + B\partial_{\xi\xi\xi}^3 \psi = 0, \tag{23}$$

where the nonlinear coefficient can be written as

$$A = \frac{3v_p}{2(1-\beta)(\beta-2)}\left[\frac{1+(1-\beta)^2}{v_p^2-2}\right],$$

and also the dispersion coefficient can be written as

$$B = \frac{1-\beta}{2v_p}\left[\frac{(v_p^2-2)^2}{2-\beta}\right].$$

The steady state solution of the K-dV equation (23) is obtained by considering a moving frame (moving with speed $u_0$) $\zeta = \xi - u_0\tau$, and applying the appropriate boundary conditions, viz. $\psi \to 0$, $d_\zeta\psi \to 0$, $d_\zeta^2\psi \to 0$ at $\zeta \to \pm\infty$. After some algebraic calculations (details can be found in Ref. [36]), one can express the steady state solitonic solution of this K-dV equation as

$$\psi = \psi_0 sech^2\left[\frac{\zeta}{\Delta}\right], \tag{24}$$

where the maximum potential amplitude or height $\psi_0$ and the width or thickness $\Delta$ of the solitonic profile, are given by

$$\psi_0 = 3u_0/A = \frac{u_0(1-\beta)(\beta-2)}{v_p}\left[\frac{2(v_p^2-2)}{1+(1-\beta)^2}\right],$$
$$\Delta = \sqrt{4B/u_0} = \sqrt{\frac{2(1-\beta)}{u_0 v_p}\left[\frac{(v_p^2-2)^2}{2-\beta}\right]}.$$

## IV. DERIVATION OF THE MKDV EQUATION AND ITS SOLUTION

We have examined the electrostatic perturbations propagating in an magnetized collisionless plasma system due to the effect of dispersion. By considering higher-order term, we have derived the mKdV equation employing the reductive perturbation method using Eqs. (1)-(3). The set of stretched coordinates [37] was introduced for the mKdV equation as

$$\xi = \epsilon(x - v_p t), \quad (25)$$
$$\tau = \epsilon^3 t, \quad (26)$$

and will use the expansion (12)

$$u_\alpha = \sum_{n=2}^{\infty} \epsilon^n u_\alpha^{(n-1)}. \quad (27)$$

The lowest order in $\epsilon$ yields the relations (17)-(20) with the help of (25)-(27). The higher orders in $\epsilon$ in yields a set of expressions which are omitted here relating the second and third order contributions to the first order potential contribution $\psi$

$$\partial_\tau \psi + C\psi^2 \partial_\xi \psi + B \partial_{\xi\xi\xi}^3 \psi = 0, \quad (28)$$

where the cubic nonlinearity coefficient $C$ reads

$$C = \frac{3(1-\beta)}{4v_p(2-\beta)} \left[ \frac{\delta v_p^2 (5v_p^2 + 8\delta\sigma)}{(v_p^2 - 2\delta\sigma)^5} + \frac{v_p^2(5v_p^2 + 8)}{(v_p^2 - 2)^3} \right]. \quad (29)$$

In order to trace the influence of different plasma parameters on the propagation of electrostatic waves in our considered plasma system, we have derived mKdV equation. The stationary solitary wave solution of standard mKdV equation is obtained by considering a frame $\zeta_m = \xi - u_0 \tau$, and the solution for solitonic profile is as follows

$$\psi(m) = \psi_m sech\left[\frac{\zeta_m}{\Delta_m}\right], \quad (30)$$

where the maximum potential amplitude or height $\psi_m$ and the width or thickness $\Delta_m$ of the solitonic profile, are given by

$$\psi_m = \sqrt{6u_0/C},$$
$$\Delta_m = \sqrt{u_0/B}.$$

## V. RESULTS AND DISCUSSION

To evaluate the characteristics of the solitary profile represented by Eqs. 23 (KdV) and 28 (mKdV), we have analyzed the maximum potential amplitude $\psi_0$ and $\psi(m)$ as well as investigated how the phase velocity $V_p$ and the ion-to-electron density ratio $\beta$ change the profile of the maximum potential perturbation. Our main results are presented below:

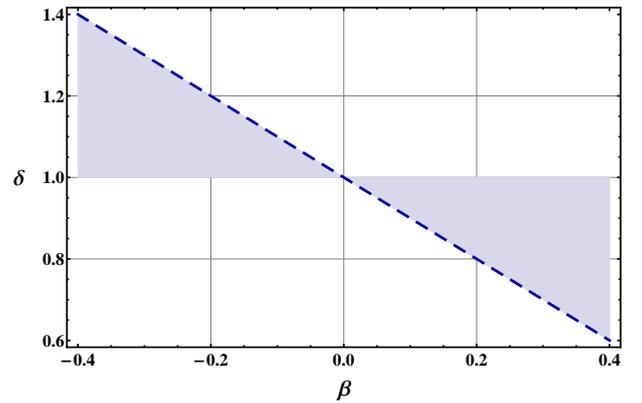

FIG. 1: $\delta$ versus $\beta$ to identify the background species.

**Determination of background species**: According to charge neutrality equation (9), $\beta$ determines the background species. If the background species is absent in that case $\beta = 0$ and ultimately to maintain quasineutrality $\delta = 1$. To present any species $\beta$ must be a non zero even negative value. For positive (negative) background species $\beta$ holds positive (negative) value. It is seen that from Fig-1, positive value of $\beta$ pair with lower value (less than unity) of $\delta$ and vice-versa to follow the charge neutrality condition.

**Variation of $V_p$ against $\beta$ for different values of $\sigma$**: In Fig-2, the dependence of the phase speed $V_p$ (given by Eqs. 21 or 22) on the background charge species $\beta$ and temperature $\sigma$ ratios is analyzed. It is found that lower value of $\beta$ (i.e., negative background species) higher value of $V_p$ with higher value of $\sigma$ and vice-versa. Here, $\sigma$ measures the temperature symmetry (when $\sigma = 1$) or asymmetry (when $\sigma < 1$ or $\sigma > 1$). So, $\sigma = 1$ reads as a critical value. Therefore, in the case of a positive background species, the higher the fixed species density (or, the more ions present in e-p-i plasmas), the slower

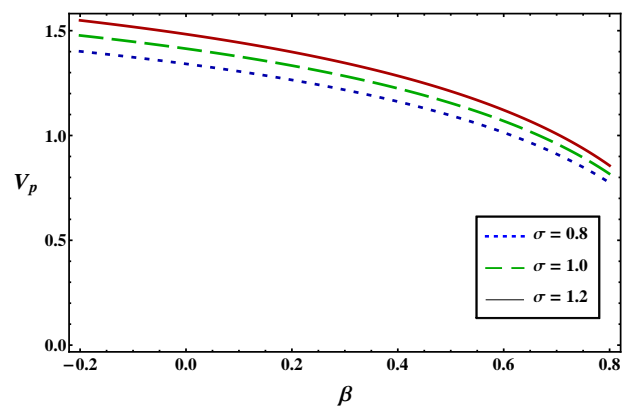

FIG. 2: $V_p$ versus $\beta$ for different values of $\sigma$.

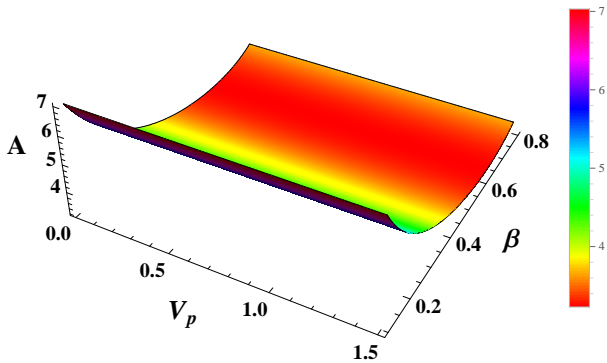

FIG. 3: $A$ versus $V_p$ and $\beta$ when $\sigma = 0.8$.

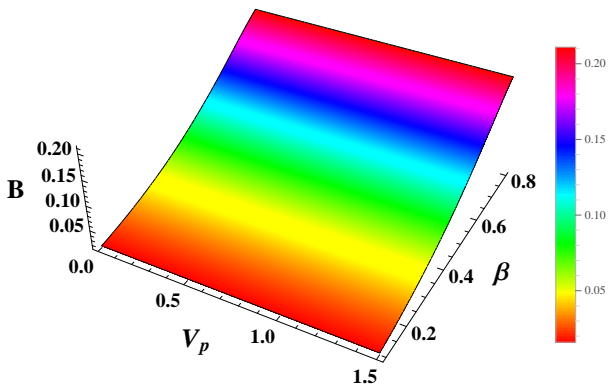

FIG. 4: $B$ versus $V_p$ and $\beta$ when $\sigma = 0.8$.

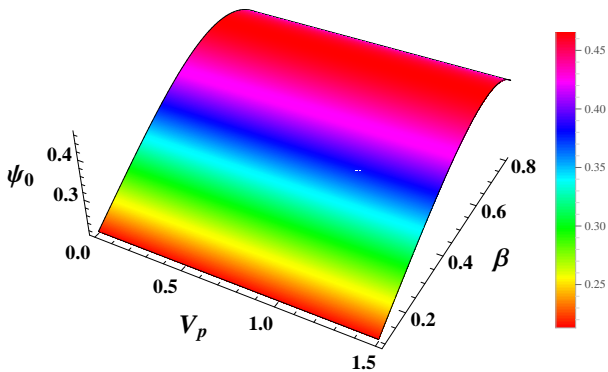

FIG. 5: $\psi_0$ versus $V_p$ and $\beta$ when $\sigma = 0.8$ and $u_0 = 0.5$.

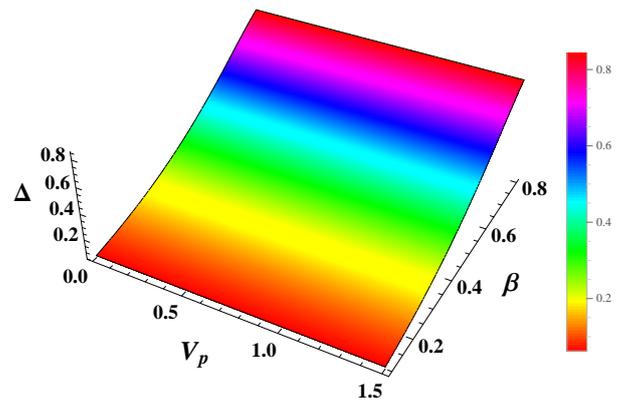

FIG. 6: $\Delta$ versus $V_p$ and $\beta$ when $\sigma = 0.8$, and $u_0 = 0.5$.

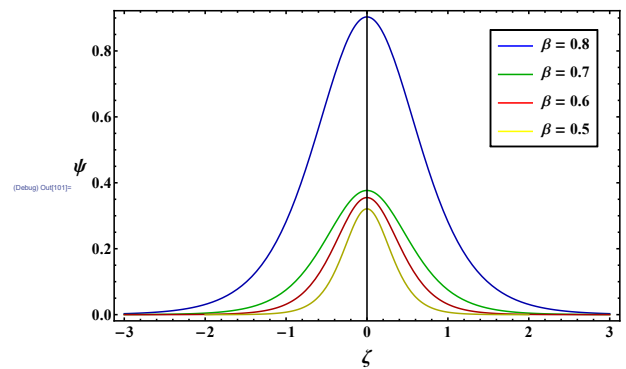

FIG. 7: Solitary wave profile for KdV when $\sigma = 0.8$ and $u_0 = 0.5$.

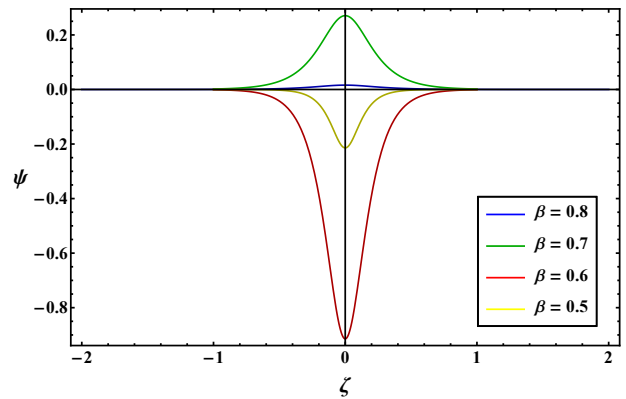

FIG. 8: Solitary wave profile for mKdV when $\sigma = 0.8$, $\delta = 0.2$ and $u_0 = 0.5$.

solitary waves will be. The opposite effect should occur, the higher negative background species concentration is.

**Variation of $A$ against $V_p$ and $\beta$**: Nonlinear coefficient $A$, with phase speed $V_p$ (invariant here), gradually decreases with increase of $\beta$ (see Fig-3).

**Variation of $B$ against $V_p$ and $\beta$**: In stark contrast, however, dispersion coefficient $B$, with phase speed $V_p$ (constant here), slowly increases with increase of $\beta$ in Fig-4.

**Variation of $\psi_0$ against $V_p$ and $\beta$**: The maximum potential $\psi_0$ increases with increase of $\beta$ (until 0.6) and after that it remains constant with the any value of $V_p$ as shown in Fig-5.

**Variation of $\Delta$ against $V_p$ and $\beta$**: The width of the solitary wave $\Delta$ increases with increase of $\beta$ with the any value of $V_p$ as shown in Fig-6.

**Variation of solitonic profile $\psi$ versus $\zeta$ for different value of $\beta$ (K-dV):** Solitonic profile is formed due to the balance between nonlinear coefficient $A$ and dispersion coefficient $B$. In our consider plasma system, solitary waves have been formed with narrow width and minimum height for the lower value of $\beta$ and it is seen that both thickness and height gradually increase with the increase of $\beta$ as shown in Fig-7.

**Variation of solitonic profile $\psi(m)$ against $\zeta_m$ for different value of $\beta$ (mK-dV):** Solitonic profile is seen to form because of the balance between nonlinear coefficient $C$ and dispersion coefficient $B$. In that case, hump shape (i.e.,+ve potential) solitary waves are formed with the value of $\beta = 0.8$ and $\beta = 0.7$ and dip shape (i.e.,-ve potential) solitary waves are formed with the value of $\beta = 0.6$ and $\beta = 0.5$. For positive potential, minimum (maximum) amplitude with minimum (maximum) wide solitary waves have been observed with $\beta = 0.8$ ($\beta = 0.7$). On the other hand, for negative potential, minimum (maximum) amplitude with minimum (maximum) wide solitary waves have been seen with $\beta = 0.5$ ($\beta = 0.6$). In stark contrast, however, the structure is distinguishable from the structure of solitary waves of KdV (shown in Fig-8). For KdV only positive potential (see Fig-7) is seen but for mKdV both positive and negative potentials (see Fig-8) are seen with same value of $\beta$.

## VI. CONCLUSION

We have investigated the propagation characteristics of solitary waves in a doped pair-ion or electron-positron-ion plasmas. Using the reductive perturbation technique, the KdV and mKdv equations are derived which governs the small amplitude nonlinear solitary waves. The effects of the plasma parameters, namely the positive ion temperature to negative ion temperature ratio $\sigma$, density ratio $\delta$, and charge background species $\beta$, are studied. It is found that the propagation characteristics of solitary waves are significantly modified by these plasma parameters. By way of conclusion, the results should be useful for understanding the properties of solitary waves that may be relevant in laboratory and space plasmas.